\magnification=\magstep1
\baselineskip=18pt
\hfuzz=6pt

$ $

\vskip 1in

\centerline{\bf Quantum principal component analysis} 

\bigskip

\centerline{Seth Lloyd$^{1,2}$, Masoud Mohseni$^3$, Patrick Rebentrost$^2$}

\centerline{1. Massachusetts Institute of Technology, department
of Mechanical Engineering}

\centerline{2. MIT, Research Laboratory for Electronics}

\centerline{3. Google Research}

\bigskip
\noindent{\it Abstract:} The usual way to reveal properties of an
unknown quantum state, given many copies of a system in that state,
is to perform measurements of different observables and
to analyze the measurement results statistically. 
Here we show that the unknown quantum state can play an
active role in its own analysis.  In particular, given multiple copies of
a quantum system with density matrix $\rho$, 
then it is possible to 
perform the unitary transformation $e^{-i\rho t}$.  
As a result, one can create quantum coherence among
different copies of the system to perform
quantum principal component analysis,
revealing the eigenvectors corresponding to the large eigenvalues
of the unknown state in time exponentially faster than
any existing algorithm.  


\vskip 1cm

Quantum tomography is the process of discovering features of
an unknown quantum state $\rho$ 
[1-2].   Quantum tomography is a widely used tool with important
practical applications in communication systems such as optical
channels, precision measurement devices such as atomic clocks,
and quantum computation.  The basic assumption of quantum tomography is
that one is given multiple copies
of $\rho$ in a $d$-dimensional Hilbert space, for example,
states of atoms in an atomic clock or inputs and outputs
of a quantum channel.   
A variety of measurement techniques allow one to
extract desired features of the state.  For example, recent developments
have shown quantum compressive sensing 
can give significant advantages 
for determining the unknown state or dynamics
of a quantum system, particularly when that state or dynamics
can be represented by sparse or low-rank matrices [3-5].  In conventional state 
tomography techniques, the state plays a passive role: it
is there to be measured.  This paper shows that the state
can play an active role in its own measurement.  In particular,
we show that multiple copies of the state $\rho$ can be
used to implement the unitary operator $e^{-i\rho t}$:  
that is, the state functions as an energy operator or
Hamiltonian, generating transformations on other states. 
First, we use this density matrix exponentiation to show
how to exponentiate non-sparse matrices in time $O(\log d)$,
an exponential speed-up over existing algorithms.
Next, we show that density matrix
exponentiation can provide significant advantages
for quantum tomography: the density matrix plays an
active role in revealing its own features.
Principal component analysis (PCA) is a method for
analyzing a positive semi-definite Hermitian matrix by
decomposing it in terms of the eigenvectors corresponding
to the matrices largest eigenvalues [6-7].  
Principal component analysis is commonly used to analyze
the covariance matrix of sampled random vectors. 
We use the fact that any positive semi-definite Hermitian matrix --
such as a density matrix -- can be represented in Gram form and 
thus as a covariance matrix of a set of vectors. 
Quantum principal component analysis (qPCA) uses multiple copies of
an unknown density matrix to construct
the eigenvectors corresponding to the large eigenvalues
of the state (the principal components)
in time $O(\log d)$, also an
exponential speed-up over existing algorithms. 
Finally, we show how quantum principal component analysis can provide
novel methods of state discrimination and cluster assignment.

Suppose that one is presented with $n$ copies of $\rho$.
A simple trick allows one to apply the unitary transformation
$e^{-i\rho t}$ to any density matrix $\sigma$ up to $n$th order in $t$.  
Note that
$$
\eqalign{ {\rm tr}_P ~ e^{-i S\Delta t} \rho \otimes \sigma ~ e^{i S\Delta t} 
&= (\cos^2 \Delta t) \sigma 
+ (\sin^2 \Delta t) \rho - i \sin \Delta t [ \rho, \sigma]\cr
&= \sigma -  i\Delta t [ \rho, \sigma] + O(\Delta t^2).\cr}  
\eqno(1)$$
Here ${\rm tr}_P$ is the partial trace over the first
variable and $S$ is the swap operator.  $S$ is a sparse
matrix and so $e^{-iS\Delta t}$ can be performed efficiently
[6-9].  Repeated application of (1) with
$n$ copies of $\rho$ allows one to construct 
$e^{-i\rho n\Delta t} \sigma ~ e^{i\rho  n  \Delta t}$.  Comparison
with the Suzuki-Trotter theory of quantum simulation [8-11]
shows that to simulate $e^{-i\rho t}$ to accuracy $\epsilon$
requires $n = O( t^2 \epsilon^{-1} |\rho-\sigma|^2) 
\leq O( t^2 \epsilon^{-1})$ steps,
where $t=n\Delta t$ and $|~|$ is the sup norm.  So simply performing repeated
infinitesimal swap operations on $\rho\otimes\sigma$ allows
us to construct the unitary operator $e^{-i\rho t}$.
The quantum matrix inversion techniques
of [12] then allow us to use multiple copies of a density matrix
$\rho$ to implement $e^{-ig(\rho)}$ efficiently for any simply
computable function $g(x)$. 


As a first application of density matrix exponentiation,
we show how to exponentiate low-rank
positive non-sparse $d$-dimensional Hamiltonians in time $O(\log d)$.
Existing methods using the higher order Suzuki-Trotter
expansion [8-11] require time $O(d\log d)$ to
exponentiate non-sparse Hamiltonians.
We want to construct $e^{-iXt}$ 
for non-sparse positive $X$, where the
sum of the eigenvalues of $X = 1$.  Write
$X= A^\dagger A$, where $A$ has columns
$\vec a_j$, not necessarily normalized to 1.
In quantum-mechanical form,  
$A = \sum_i |\vec a_i| |a_i\rangle
\langle e_i|$, where $|e_i\rangle$ is an orthonormal basis,
and the $|a_i\rangle$ are normalized to 1.   
Assume that we have quantum access to the columns $|a_i\rangle$
of $A$ and to their norms $|\vec a_i|$.  That is, we have a
quantum computer or quantum random access memory
(qRAM) [13-15] that takes $|i\rangle|0\rangle|0\rangle \rightarrow
|i\rangle|a_i\rangle | |\vec a_i| \rangle $.  Quantum access to
vectors and norms allows us to construct the state
$ \sum_i |\vec a_i| |e_i\rangle |a_i\rangle$ [19]: 
the density matrix for the first register is
exactly $X$.  Using  
$n = O(t^2\epsilon^{-1})$ copies of $X$ allows us to    
implement $e^{-iX t}$ to accuracy $\epsilon$ in
time $O( n \log d)$.  

Note that density matrix exponentiation is most effective
when some of the eigenvalues of $\rho$ are large.   If all the eigenvalues
are of size $O(1/d)$ then we require time $t = O(d)$ to generate
a transformation that rotates the input state $\sigma$ to
an orthogonal state.  By contrast, if the density matrix 
matrix is dominated by a few large
eigenvalues -- that is, when the matrix is well represented
by its principal components -- then the method works
well (the accuracy will be analyzed below).  
In this case, there exists a subspace
of dimension $R << d$ such that the projection of 
$\rho$ onto this subspace is close to $\rho$:
$\| \rho - P\rho P\|_1 \leq \epsilon$, where $P$ is the projector
onto the subspace.  
When the matrix is of low rank, the projection is
exact.  Current techniques for
matrix exponentiation are efficient when the matrix to
be exponentiated is sparse [9-10].  The construction here
shows that non-sparse but low-rank matrices can also
be exponentiated efficiently. 

Density matrix exponentiation now allows us to apply
the quantum phase algorithm to find the 
eigenvectors and eigenvalues of an unknown density matrix.
If we have $n$ copies of $\rho$, use the ability to
apply $e^{-i\rho t}$ to perform the quantum phase
algorithm [1].  In particular, the quantum phase
algorithm uses conditional applications of $e^{-i\rho t}$
for varying times $t$ to take any initial state $|\psi\rangle |0\rangle$
to $\sum_i \psi_i |\chi_i\rangle |\tilde r_i\rangle$,
where $|\chi_i\rangle$ are the eigenvectors of $\rho$
and $\tilde r_i$ are estimates of the corresponding    
eigenvalues.  Using the improved phase-estimation
techniques of [12] yields the eigenvectors
and eigenvalues to accuracy $\epsilon$ by applying
the quantum phase algorithm for time $t = O(\epsilon^{-1})$,
and so requires
$ n = O(1/\epsilon^3)$ copies of the state $\rho$.  
Using $\rho$ itself as the initial
state, the quantum phase algorithm yields the state
$$ \sum_i r_i |\chi_i\rangle\langle \chi_i| \otimes 
|\tilde r_i\rangle \langle \tilde r_i|.\eqno(2)$$
Sampling from this state allows us to reveal features
of the eigenvectors and eigenvalues of $\rho$.  

As above, quantum self-tomography is particularly useful when 
$\rho$ can be decomposed accurately into its principal components.
For example, if the rank
$R$ of $\rho$ is small,  only $R$ eigenvectors
and eigenvalues are represented in the eigenvector/eigenvalue
decomposition (2), and the average size of $r_i$ is $1/R$.  
Using $mn$ copies of $\rho$ we obtain $m$ copies
of the decomposition (2), where the $i$'th eigenvector/eigenvalue
appears $r_i m$ times.  The features of the $i$'th eigenstate
can then be determined by measuring
the expectation value $\langle \chi_i| M |\chi_i\rangle$
of the eigenvector with eigenvalue $r_i$ for arbitrary
Hermitian $M$.   Note that we are no longer restricted
to evaluating only expectation values of sparse matrices. 
As long as the trace of $\rho$ is dominated by a few large
eigenvalues, then quantum self-tomography can be used to
perform principal component analysis on
the unknown density matrix $\rho$. 
For example, suppose that the density matrix corresponds
to the covariance matrix of a
set of data vectors $|a_i\rangle$ that can be generated
in quantum parallel using the oracle above. 
Quantum principal component
analysis then allows us to find and to work with the directions 
in the data space that have the largest variance.

State self-tomography can be extended to
quantum process self-tomography by using the Choi-Jamiolkowski state 
$(1/d) \sum_{ij} |i\rangle\langle j|\otimes {\cal S}( |i\rangle\langle j|)$
for a completely positive map ${\cal S}$ [16].
For quantum channel tomography, for example, the Choi-Jamiolkowski
state is obtained by sending half of a fully entangled
quantum state down the channel.  Quantum principal component analysis
then be used to construct the eigenvectors corresponding to
the dominant eigenvalues of this state: the resulting
spectral decomposition in turn encapsulates many of
the most important properties of the channel [17]. 

Comparing quantum self-tomography to quantum compressive sensing [3-5],
we see that self-tomography holds several advantages in terms of scaling.
Self-tomography is not confined to sparse matrices; 
most importantly, self-tomography reveals eigenvectors and
eigenvalues in time $O(R\log d)$ compared
with $O(Rd\log d)$ for compressive tomography [3].
Of course, quantum self-tomography cannot reveal all the
$d^2$ entries of $\rho$ in time $R\log d$: but it can 
present the eigenvectors of $\rho$ in quantum form so
that their properties can be tested.

Quantum self-tomography shows that the density matrix exponentiation
presented here is time-optimal.  One might think, in analogy
to the use of higher order terms in the Suzuki-Trotter
expansion for exponentiation of sparse matrices [8-11], that 
could be possible to reduce the number of copies required
to perform density matrix exponentiation
to accuracy $\epsilon$ over time $t$ to $O(t/\epsilon)$.
If one could do this, however, the self-tomography
algorithm just given would allow us to find the
eigenvalues of an unknown density matrix to accuracy
$\epsilon = O(1/n)$ using $n$ copies of the matrix.
Even if the eigenbasis of the density matrix is known,
however, sampling $n$ copies of the density
matrix only allows one to determine the eigenvalues
of the matrix to accuracy $O(1/\sqrt n)$ [17].
Quantum self-tomography can be compared to group representation
based methods for estimating the spectrum of a density matrix [18]
(with the difference that quantum self-tomography also reveals
the eigenvectors).

Quantum principal component analysis can also be useful in state discrimination
and assignment.  For example, suppose that we can sample
from two sets of $m$ states, the first set
 $\{|\phi_i\rangle\}$ characterized by a density matrix 
$\rho = (1/m) \sum_i|\phi_i \rangle\langle \phi_i|$, and the
second set $\{ |\psi_i\rangle\}$ characterized 
by a density matrix $\sigma = (1/m) \sum_i |\psi_i\rangle
\langle \psi_i|$.  Now we are given a new state $|\chi\rangle$.
Our job is to assign the state to one set or the other.
Density matrix exponentiation and quantum phase estimation
then allow us to decompose $|\chi\rangle$ in terms of
the eigenvectors and eigenvalues of the $\rho -\sigma$:
$$|\chi\rangle|0\rangle \rightarrow 
\sum_j \chi_j | \xi_j\rangle |x_j\rangle, \eqno(3)$$
where $|\xi_j\rangle$ are the eigenvectors of 
$\rho-\sigma$ and $x_j$ are the corresponding eigenvalues.
Measure the eigenvalue register,
and assign $|\chi\rangle$ to the first set
if the eigenvalue is positive and to the second set if it is
negative.  If $|\chi\rangle$  is selected from one of the two
sets, this procedure is simply minimum error
state discrimination [1], but with a bonus.  The magnitude of the measured
eigenvalue is a measure of the confidence of the set
assignment measurement: larger magnitude eigenvalues correspond
to higher confidence in the assignment, 
and magnitude 1 correponds to certainty -- in this case 
 $|\xi\rangle$ is orthogonal to all the members of one
of the sets.  If $|\chi\rangle$ is some other vector, then
the method provides a method for supervised learning and
cluster assignment [6-7, 19]: the two sets are training sets and the
vector is assigned to the set of vectors to which it is
more similar.

\bigskip\noindent{\it Discussion:} Density matrix exponentiation
represents a powerful tool for analyzing the properties of unknown
density matrices.   The ability to use $n$ copies of
$\rho$ to apply the unitary operator $e^{-i\rho t}$ allows
us to exponentiate non-sparse $d$-dimensional matrices to
accuracy $\epsilon = O(1/\sqrt n)$ in
time $O(\log d)$, and to perform quantum self-tomography
to construct the eigenvectors and eigenvalues of $\rho$
in time $O(R\log d)$.  In such quantum self analysis, the
density matrix becomes an active participant in 
the task of revealing its hidden features.  

Like quantum matrix inversion [12], quantum principal component analysis 
maps a classical procedure that takes time polynomial
in the dimension of a system to a quantum procedure
that takes time polynomial in the logarithm of the dimension.
This exponential compression means that quantum principal
component analysis 
can only reveal a fraction of the full information required
to describe the system.  That particular fraction of information can
be very useful, however, as the ability of density matrix
inversion to reconstruct its principal components shows.

We anticipate that quantum principal componetn can play a key role
in a variety of quantum algorithms and measurement applications.
As the example of quantum cluster assignment shows, quantum
self analysis could be useful for 
speeding up to machine learning problems such as clustering
and pattern recognition [6-7, 19].  The ability to 
identify the largest eigenvalues of a matrix together
with the corresponding eigenvectors 
is potentially useful for the representation and 
analysis of large amounts of high-dimensional data.

\vfill
\noindent{\it Acknowledgments:} This work was supported by DARPA,
Google, and Jeffrey Epstein.  The authors thank Andrew Childs and
Robin Kothari for helpful discussions.
\vfil\eject
\noindent{\it References:}

\bigskip

\smallskip\noindent [1] M.A. Nielsen, I.L. Chuang, {\it Quantum computation and
quantum information}, Cambridge University Press, Cambridge (2000).

\smallskip\noindent [2] M. Mohseni, A.T. Rezakhani, D.A. Lidar,
{\it Phys. Rev. A} {\bf 77}(3), 032322 (2008).

\smallskip\noindent [3] D. Gross, Y.-K. Liu, 
S.T. Flammia, S. Becker, J. Eisert,
{\it Phys. Rev. Lett.} {\bf 105}, 150401 (2010);
arXiv: 0909.3304. 

\smallskip\noindent [4] A. Shabani, R. L. Kosut, M. Mohseni, H. Rabitz, 
M. A. Broome, M.P. Almeida, A. Fedrizzi, A. G. White,
{\it Phys. Rev. Lett.} {\bf 106}, 100401 (2011);
arXiv: 0910.5498. 

\smallskip\noindent [5]  A. Shabani, M. Mohseni, S. Lloyd, 
R.L. Kosut, H. Rabitz,
{\it Phys. Rev. A} {\bf 84}, 012107 (2011);
arXiv: 1002.1330. 

\smallskip\noindent [6] Christopher M. Bishop, 
{\it Pattern Recognition and Machine Learning,} Springer, New York,
2006.

\smallskip\noindent [7]
Kevin Patrick Murphy, 
{\it Machine Learning: a Probabilistic Perspective,} MIT Press, 
Cambridge, 2012.

\smallskip\noindent [8] S. Lloyd, {\it Science} {\bf 273}, 1073-1078 (1996).

\smallskip\noindent [9] D. Aharonov, A. Ta-Shma, {\it Proc. 35th Annual
ACM Symp. on Theory of Computing}, 20-29 (2003).

\smallskip\noindent [10] D.W. Berry, G. Ahokas, R. Cleve, and B. C. Sanders, 
{\it Comm. Math. Phys.} {\bf 270}, 359-371 (2007);
arXiv: quant-ph/0508139.

\smallskip\noindent [11]  N. Wiebe, D.W. Berry, P. Hoyer, B.C. Sanders,
{\it J. Phys. A: Math. Theor.} {\bf 43}, 065203 (2010);
arXiv: 0812.0562. 


\smallskip\noindent [12] A.W. Harrow, A. Hassidim, S. Lloyd,
{\it Phys. Rev. Lett.} {\bf 103}, 150502 (2009);
arXiv: 0811.3171.

\smallskip\noindent [13] V. Giovannetti,
S. Lloyd, L. Maccone,  {\it Phys.Rev.Lett.} {\bf 100},
160501 (2008); arXiv: 0708.1879.

\smallskip\noindent [14] V. Giovannetti,
S. Lloyd, L. Maccone, {\it Phys.Rev.A} {\bf 78},
052310 (2008); arXiv: 0807.4994.

\smallskip\noindent [15] F. De Martini, V. Giovannetti, S. Lloyd, L. Maccone,
E. Nagali, L. Sansoni, F. Sciarrino, {\it Phys. Rev. A } {\bf 80},
010302(R) (2009); arXiv: 0902.0222.

\smallskip\noindent [16] M.-D. Choi, {\it Lin. Alg. Appl.}
{\bf 10}, 285-290 (1975). 

\smallskip\noindent [17] V. Giovannetti, S. Lloyd, L. Maccone,
{\it Phys. Rev. Lett.} {\bf 96}, 010401 (2006); arxiv: 0509179.

\smallskip\noindent [18] M. Keyl, R.F. Werner, {\it Phys. Rev. A}
{\bf 64}, 052311 (2001).

\smallskip\noindent [19] S. Lloyd, M. Mohseni, P. Rebentrost, 
`Quantum algorithms for supervised and unsupervised machine learning,' 
arXiv: 1307.0411; 
P. Rebentrost, S. Lloyd, M. Mohseni, 
`Quantum support vector machine for big feature and big
data classification,' arXiv: 1307.0471.

\vfill\eject\end